\documentclass[aps,prl,amsmath,amssymb,superscriptaddress,preprintnumbers,twocolumn,
 amsmath,amssymb,
 aps,
]{revtex4-2}

\usepackage{graphicx}% Include figure files
\usepackage{dcolumn}% Align table columns on decimal point
\usepackage{bm}% bold math
\usepackage{cancel}

\usepackage[usenames, dvipsnames]{xcolor}
\usepackage{tikz}
\usetikzlibrary{shapes}
\usepackage[co mpat=1.1.0]{tikz-feynman}
\usepackage{hyperref}% add hypertext capabilities
\begin{document}

%\preprint{.......}

\title{Exact Results for Scaling Dimensions of Neutral Operators in scalar CFTs
}

\author{Oleg Antipin}%
\email{oantipin@irb.hr}
\affiliation{ Rudjer Boskovic Institute,
  Division of Theoretical Physics,
  Bijeni\v cka 54, 10000 Zagreb, Croatia}
  
\author{Jahmall Bersini}
\email{jahmall.bersini@ipmu.jp}
\affiliation{Kavli IPMU (WPI), UTIAS, The University of Tokyo, Kashiwa, Chiba 277-8583, Japan}

\author{Francesco Sannino}
\email{sannino@qtc.sdu.dk}
\affiliation{Quantum Theory Center ($\hslash$QTC) at IMADA \& D-IAS, Southern Denmark Univ., Campusvej 55, 5230 Odense M, Denmark}
\affiliation{Dept. of Physics E. Pancini, Universit\`a di Napoli Federico II, via Cintia, 80126 Napoli, Italy}
\affiliation{INFN sezione di Napoli, via Cintia, 80126 Napoli, Italy}
\affiliation{Scuola Superiore Meridionale, Largo S. Marcellino, 10, 80138 Napoli, Italy}

\begin{abstract}
We determine the scaling dimension $\Delta_n$ for the class of composite operators $\phi^n$ in the $\lambda \phi^4$ theory in $d=4-\epsilon$ taking the double scaling limit  $n\rightarrow \infty$ and $\lambda \rightarrow 0$ with fixed $\lambda n$  via a semiclassical approach. Our results resum the leading power of $n$ at any loop order.  In the small $\lambda  n$ regime we reproduce the known diagrammatic results and predict the infinite series of higher-order terms. For intermediate values of $\lambda n$  we find that $\Delta_n/n$ increases monotonically approaching a $(\lambda n)^{1/3}$ behavior 
in the $\lambda n \to \infty$ limit. We further generalize our results to neutral operators in the  $\phi^4$ in $d=4-\epsilon$, $\phi^3$  in $d=6-\epsilon$, and $\phi^6$ in $d=3-\epsilon$ theories with $O(N)$ symmetry.
 \end{abstract}

 \maketitle

Critical behavior of quantum field theories plays a crucial role in our understanding of phase transitions in Nature across all realms of physics from condensed matter to high energy physics and cosmology. 
Such behavior is encoded in scaling exponents associated with correlation functions of different operators of the underlying conformal field theory (CFT) employed to describe a given physical process. 

Since the pioneering work of Wilson \cite{Wilson:1973jj, Wilson:1974mb}, scalar field theories have been used as primary models to unveil universal behavior in phase transitions. Despite the fifty-year-long effort to solve these models much is still left to be understood. Perhaps one of the greatest challenges is the investigation of composite operators. For these, perturbation theory, very quickly shows its limitations. Beyond perturbation theory a number of methodologies have been employed from the use of larger symmetries, such as supersymmetry, to large charge \cite{Hellerman:2015nra} and/or spin \cite{Komargodski:2012ek} expansions, bootstrap \cite{Rattazzi:2008pe}, and numerical approaches.

In this work, we develop a novel way to determine the scaling dimensions $\Delta_n$ of neutral operators, schematically $\phi^n$, in scalar CFTs in the double-scaling limit of large $n$, small self coupling $\lambda$, and generic $\lambda n$. We will employ the methodology to tackle various scalar CFTs living in different space-time dimensions $d$. We discover that, for all the models, the large $ \lambda n$ behavior is of the type: $\Delta_n \propto n^{d/(d-1)}  $. We therefore conjecture that this leading large $n$ scaling holds non-perturbatively away from the small $\lambda$ limit. Interestingly, this behavior mimics the one found for large charge operators with charge $n$ \cite{Hellerman:2015nra, Badel:2019oxl}.

 \vskip .2cm
\centerline{\bf Methodology and $\phi^4$ theory in $d=4-\epsilon$}
 \vskip .1cm
A cornerstone example of CFT is the critical $\lambda \phi^4$ theory in $d=4-\epsilon$ dimensions.  We use this model to introduce the  semiclassical framework to determine the scaling dimensions $\Delta_n$ controlling the critical behavior of the correlator 
\begin{equation} 
\langle \phi^n(x_f) \phi^n(x_i) \rangle \ = \frac{1}{\rvert x_f - x_i \rvert^{2\Delta_{n}}} \ ,
\end{equation} 
at the Wilson-Fisher infrared fixed point stemming from the Lagrangian below
\begin{equation}
    \mathcal{L} =\frac{1}{2}(\partial \phi)^2 - \frac{\lambda}{4} \phi^4 \ .
\end{equation}
 The two-loop fixed-point coupling value is 
\begin{equation} \label{FP}
\lambda^* =\frac{8 \pi ^2 }{9}  \epsilon + \frac{136 \pi ^2}{243}  \epsilon ^2+ \mathcal{O}\left(\epsilon^3\right)\ .
\end{equation}
For general $n$, we compute the three-loop value of $\Delta_n$ and obtain 
\begin{align} \label{check}
    \Delta_n &=n \left(1-\frac{\epsilon }{2}\right)+\frac{n}{6} (n-1) \epsilon 
    -\frac{ \epsilon ^2 }{324}  \left(17 n^3-67 n^2+47 n \right)   \nonumber \\ & + \frac{n}{34992} \Big(1125 n^3-7433 n^2+15034 n-8399 \nonumber \\ & +2592 (n-3) (n-1) \zeta (3) \Big)\epsilon ^3+\mathcal{O}\left(\epsilon^4\right) \ .  
\end{align}
The two-loop result has been previously determined in \cite{Derkachov:1997gc}.
Determining  $\Delta_n$, at arbitrary orders in perturbation theory, is an involved task. Within the path integral formalism, calculating $\Delta_n$ amounts to perform the following functional integration
% \begin{equation} 
% \langle \phi^n(x_f) \phi^n(x_i) \rangle \ = \frac{1}{\rvert x_f - x_i \rvert^{2\Delta_{n}}}
% \end{equation} 
\begin{equation} 
\langle \phi^n(x_f) \phi^n(x_i) \rangle \ = \int \mathcal{D} \phi  \ \phi^n(x_f) \phi^n(x_i) e^{i \int d^d x \mathcal{L}} \,.
\end{equation} 
Upon exponentiating the field insertion and rescaling the field as $\phi \to \sqrt{n} \phi$  we observe that  $n$  becomes a counting parameter. As a consequence, the above correlator can be estimated semiclassically around the saddle point of the following action:  
\begin{equation}
     n \left[ \int d^d x \left(  \frac{1}{2}(\partial \phi)^2 - \frac{\lambda n}{4} \phi^4 \right)  - i  \, (\log \phi(x_f) + \log \phi (x_i) ) \right]\ .
\end{equation}
Further employing  the double scaling limit  $n\rightarrow \infty$, $\lambda \rightarrow 0$ with fixed $\lambda n$  yields the following expansion for $\Delta_n$
\begin{equation} \label{dsl}
    \Delta_n = n \sum_{i=0} \frac{C_i(\lambda n)}{n^i} \ ,
\end{equation}
where the coefficients $C_i$ arise from the $i$-th order of the semiclassical expansion. A similar approach has been used to determine multiparticle scattering amplitudes and decay rates \cite{Brown:1992ay, Son:1995wz}, and also to compute scaling dimensions of large charge composite operators in theories with continuous symmetries \cite{Hellerman:2015nra, Badel:2019oxl}.

For CFTs the computation is efficiently performed by conformal mapping flat space into a cylinder $\mathbb{R}\times S^{d-1}$ with unit radius. According to  Cardy's state-operator correspondence \cite{Cardy:1984epx} a given scaling dimension becomes the energy on the cylinder of the corresponding state. 
On the cylinder the Lagrangian reads
\begin{equation}
    \mathcal{L} =\frac{1}{2}(\partial \phi)^2 -\frac{(d-2)R}{(d-1)8}  \phi^2-\frac{\lambda}{4} \phi^4 \ ,
\end{equation}
with the Ricci curvature  $R=(d-1)(d-2)$. In this work, we will determine the leading coefficient $C_0$ of the semiclassical expansion which is given by the classical energy on the cylinder. To the leading order in the expansion \eqref{dsl}, one can set $d=4$ since the classical theory is scale invariant.

To compute the energy on the cylinder we solve the following time-dependent equation of motion (EOM)
\begin{equation}
\frac{d^2\phi}{dt^2} + \phi + \lambda \phi^3 = 0 \ ,
\label{eom}
\end{equation}
assuming a spatially homogeneous field configuration supplemented by the Bohr-Sommerfeld condition
\begin{equation}
2\pi^2\int_{0}^{{\cal T}} \left(\frac{d\phi}{dt} \right)^2 \,dt = 2\pi n \ ,  
\end{equation}
needed to select the appropriate state in the theory. Here ${\cal T}$ is the period of the solution which depends on the product $\lambda n$. 
The leading order of the semiclassical expansion $C_0$ can now be obtained by evaluating the energy on the solution of the equation of motion. This procedure yields
\begin{equation}
   \frac{n}{2\pi^2}\,C_0 =T_{00} = \frac{1}{2} \left(\frac{\partial \phi}{\partial t} \right)^2 + \frac{1}{2} \phi^2 +\frac{ \lambda}{4}  \phi^4 \ ,
\end{equation}
with $T_{\mu \nu}$ the stress-energy tensor of the theory and the $2\pi^2$ factor being the volume of $S^3$. $C_0$ resums the terms with the leading power of $n$ at any loop order. 
The general solution found in \cite{Sanchez} is
\begin{equation}
\phi(t) = \sqrt{n} \, x_0 \, \text{cn}(\omega t|m) \ ,
\end{equation}
where $\text{cn}(\omega t|m) $ denotes the Jacobi elliptic function with the frequency and the initial position given by 
\begin{equation}
    x_0 = \sqrt{\frac{2 m}{\lambda n (1-2 m)}} \,, \qquad  \omega = \frac{1}{\sqrt{1-2 m}} \ .
\end{equation}
The corresponding energy yields the leading order in the semiclassical expansion for  $\Delta_n$  which reads
\begin{equation}
\label{C0}
 C_{0} (\lambda n)  =  \frac{2\pi^2  m \,(1-m) } {\lambda n \,(1-2 m)^2} \ ,
\end{equation}
where $ 0 \leq m \leq 1/2$ is a function of $\lambda n$ which is determined by solving the Bohr-Sommerfeld condition with  ${\cal T} = 4 \mathcal{K}/\omega$ where $\mathcal{K}(m)$ is the complete elliptic integral of the first kind. Naturally ${\cal T}$ is the period of $\text{cn}(\omega t|m) $ and therefore of the solution $\phi(t)$. We obtain
%\begin{widetext}
\begin{align}
\label{mlambdan}
    \lambda n &= \frac{8 \pi }{3 (1-2 m)^{3/2}} \left[ (2m-1)\mathcal{E}(m) + (1-m)\mathcal{K}(m) \right] \ .
\end{align}
%\end{widetext}
Here  $\mathcal{E}$ denotes the complete elliptic integral of the second kind. Equation \eqref{C0} supplemented by \eqref{mlambdan}  constitutes our main result. To build some intuition let us consider first the limit $m\rightarrow 0$ where one has the known solution of the harmonic oscillator with unit frequency. This is the trivial free-field theory limit $\lambda =0$ discussed in \cite{Cuomo:2024fuy} for which $\Delta_n=n$. This result is obtained by noting that for $\lambda n \ll 1$ one has $\displaystyle{m \sim \frac{\lambda n}{2\pi^2}}$. In fact, in this regime, the solution to the EOM reduces to 
\begin{equation}
    \phi(t) =\frac{\sqrt{n}}{\pi} \cos(t) + \mathcal{O}\left(\lambda n \right) \,,
\end{equation}
and has period ${\cal{T}} = 2\pi$. The $\lambda n \ll 1$ limit maps into ordinary perturbation theory and will be discussed later in the text. 

\begin{figure}[t!]
\centering
\includegraphics[width=0.4\textwidth]{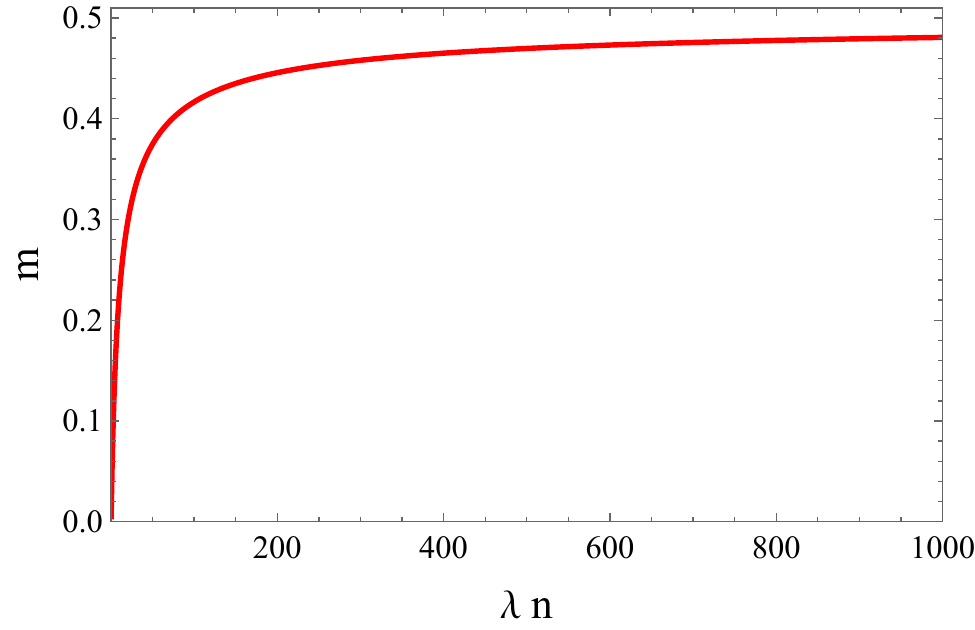} \includegraphics[width=0.4\textwidth]{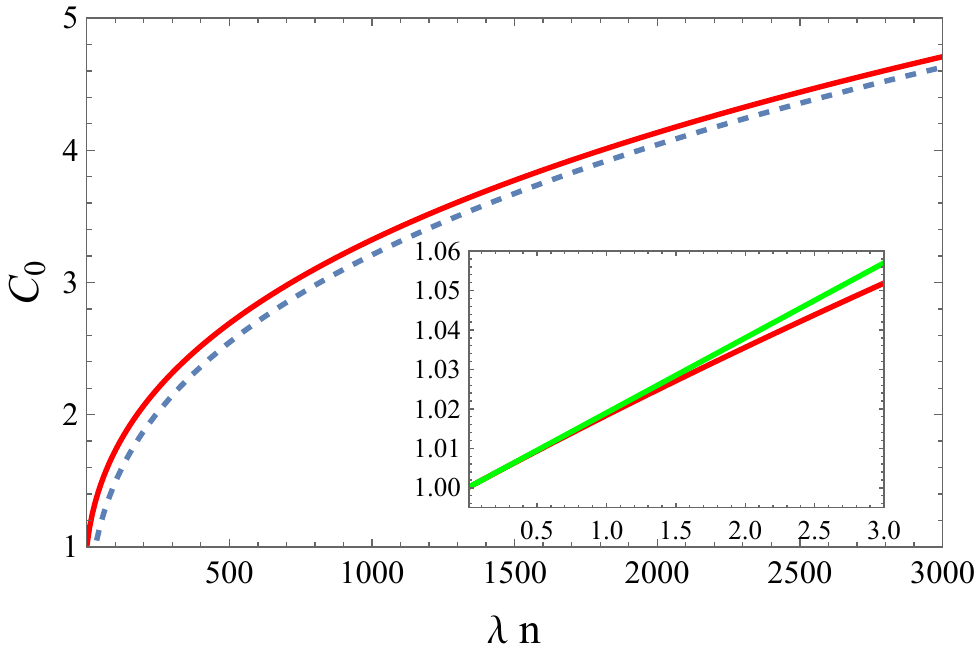} 
	\caption{The parameter $m$ (\emph{Top}) and the leading order scaling dimension $C_0$  (\emph{Bottom}) as a function of $\lambda n$. The dashed line denotes the leading large $\lambda n$ behavior of $C_0$ given by Eq. \eqref{Largeorder}. The inset plot shows a detail of $C_0$ in the small $\lambda n$ regime along with the one-loop approximation (in green).} 
	\label{mlambdanfig}
\end{figure}

When the anharmonic term dominates, for $\lambda n  \gg 1$, we observe that $m$ approaches $1/2$ from below, and for $m=1/2$, one obtains the interesting solution of the pure quartic anharmonic oscillator. The transcendental equation in \eqref{mlambdan} can be solved numerically for any $\lambda n $ with its solution given graphically in Fig.~\ref{mlambdanfig}. Here it is clear that $m$ grows monotonically with $\lambda n$ achieving asymptotically the value $m=1/2$. In the other panel of Fig.~~\ref{mlambdanfig} we plot the leading order value for $\Delta_n/n$ in the semiclassical expansion. Its behavior can be summarized as follows:

\begin{itemize}
\item[i)]{In the $\lambda n \to \infty$  limit $m$ reads 
% \begin{equation}
%     m = \frac{1}{2} -2  \left(\frac{ 2\sqrt{\pi}}{3\,\lambda n}\right)^{2/3} \Gamma \left[ \frac{5}{4}\right]^{4/3} + \mathcal{O}\left((\lambda n)^{-4/3}
% \right) \nonumber 
% \end{equation}
% leading to $\Delta_n$  asymptotically approaching 
\begin{equation}
    m = \frac{1}{2} -\pi  \left(\frac{\Gamma \left(\frac{1}{4}\right)}{6 \Gamma \left(\frac{3}{4}\right)}\right)^{2/3} \left(\frac{1}{\lambda n}\right)^{2/3}  + \mathcal{O}\left((\lambda n)^{-4/3}
\right) \,, 
\end{equation}
leading to 
\begin{equation} \label{Largeorder}
\Delta_n =\left(\frac{3\Gamma \left(\frac{3}{4}\right)}{2^{5/4}\Gamma \left(\frac{1}{4}\right)}\right)^{4/3} \lambda^{1/3} n^{4/3}  + \mathcal{O}\left(n^{2/3} \lambda^{-1/3} \right) \ .
\end{equation} We deduce a leading $n^{4/3}$ dependence in the large $\lambda n$ limit. This is the same scaling observed for the scaling dimension of large charge operators, with their charge playing the role of $n$ \cite{Hellerman:2015nra, Badel:2019oxl}.}

\item[ii)]{For intermediate $\lambda n$ we observe a smooth increase with $\lambda n$.  }

\item[iii)]{
 At small $\lambda n$ we recover both the free field theory limit as well as the conventional diagrammatic expansion as we will detail momentarily. }
\end{itemize}
The loop expansion is obtained by expanding Eq. \eqref{C0} around $\lambda n = 0$. We adopt the notation $C_0 = \sum_{k=0} a_k \left(\frac{\lambda n}{\pi^2} \right)^k$ and list the first $13$ coefficients below 
% \begin{equation}
%   C_0 =1+\frac{3 \lambda  n}{16 \pi ^2}  -\frac{17 \lambda ^2 n^2}{256 \pi ^4} + \mathcal{O}\left(\lambda^3 n^3 \right)\ .
%   \label{C0coeff}
% \end{equation}
\begin{align}
 & a_0 =1 \,, \quad a_1 =   \frac{3}{16} \,, \quad a_2 =  -\frac{17 }{256}   \,, \quad a_3 = \frac{375}{8192} \,, \nonumber \\ & a_4 = -\frac{10689}{262144} \,, \quad a_5 = \frac{87549}{2097152} \,, \quad a_6 = -\frac{3132399}{67108864} \,, \nonumber \\ & a_7 = \frac{238225977}{4294967296}\,, \quad a_8 = -\frac{18945961925}{274877906944} \,,  \nonumber \\ & a_9 = \frac{194904116847}{2199023255552} \,, \quad a_{10} = -\frac{8240234242929}{70368744177664} \,, \nonumber \\ & a_{11} = \frac{11128512976035}{70368744177664} \,, \quad a_{12} = -\frac{15671733036451359}{72057594037927936}
  \,,  \nonumber \\ & a_{13} = \frac{87535900033269525}{288230376151711744} \ .   
  %\\ & a_{14}= -\frac{7925536921177219335}{18446744073709551616} \,, \nonumber \\ & a_{15} = \frac{1451374598407735283589}{2361183241434822606848}
\end{align}
Inserting the Wilson-Fisher fixed point value Eq. \eqref{FP}, the above agrees with the diagrammatic result in Eq. \eqref{check}. Note that the $a_i$ coefficients reproduce also the known anomalous dimension of the $\phi^n$ operator in $d=4$. In fact, the small $\lambda n$ expansion of $C_0$ yields results valid also away from the fixed point. Similarly, one can now predict the terms with the leading power of $n$ to arbitrarily high loop orders.
\vskip .2cm
\centerline{\bf The $O(N)$ model}
 \vskip .1cm
We now extend our analysis 
 to  the $O(N)$ model in $d=4-\epsilon$ dimensions. The fixed point value to two-loop order is
\begin{equation} \label{ONFP}
\lambda^* =\frac{8 \pi ^2 }{(N+8)}  \epsilon + \frac{24 \pi ^2(3N+14)}{(N+8)^3}  \epsilon ^2+ \mathcal{O}\left(\epsilon^3\right) \ ,
\end{equation}
and 
% \small
% \begin{align}
%    & \Delta_n =n \left(1-\frac{\epsilon}{2} \right)+\frac{n}{2(N+8)} (3n+N-4) \epsilon -\frac{17 n^3\epsilon ^2}{4(N+8)^2}    \nonumber +\\ &
%     (n \epsilon)^2\left[\frac{604+(10-11 N)N}{4(N+8)^3} - \frac{576-N(118+35 N)}{4n(N+8)^3}\right]+\mathcal{O}(\epsilon^3) \,, 
%     \textbf{\label{ON}}
% \end{align}
% \normalsize
\small
\begin{align}
   & \Delta_n =n \left(1-\frac{\epsilon}{2} \right)+\frac{n}{2(N+8)} (3n+N-4) \epsilon - \Bigg[\frac{17}{4(N+8)^2} n^3    \nonumber \\ &
  -\frac{604+(10-11 N)N}{4(N+8)^3}n^2 + \frac{576-N(118+35 N)}{4(N+8)^3}n\Bigg]\epsilon ^2+\mathcal{O}(\epsilon^3) \,, 
    \textbf{\label{ON}}
\end{align}
\normalsize
is the two-loop value of $\Delta_n$ for the singlet operator $(\phi_a \phi_a)^{n/2}$ with $a=1, \dots, N$ \cite{Derkachov:1997gc}. 

By recognizing that by an $O(N)$ rotation the modulus coincides with one of the scalar field directions 
the EOM reduces to the one of the $N=1$ case discussed above. The dependence on $N$, to the leading order in $1/n$, appears via the fixed point value of the coupling shown above. 
Therefore $C_0$ will be again given by Eq. \eqref{C0} and Eq. \eqref{mlambdan} with $\lambda$ the $O(N)$ fixed point coupling in Eq. \eqref{ONFP}.

 \vskip .2cm
\centerline{\bf The $\phi^3$ theory in $d=6-\epsilon$}
 \vskip .1cm
To illustrate the generality of the approach we further consider the $O(N)$ $\phi^3$ theory in $d=6-\epsilon$ dimensions. Besides being a textbook example of quantum field theory, the $N=1$ case is physically relevant being related to the Lee-Yang edge singularity and percolation problems \cite{Fisher:1978pf, deAlcantaraBonfim:1981sy}. For large enough $N$ the theory features a perturbative infrared Wilson-Fisher fixed point that is believed to provide a UV completion to the quartic $O(N)$ model in more than four dimensions \cite{Fei:2014yja}. Intriguingly, it has also been proposed that the $d$ dimensional $O(N)$ CFT has a dual holographic description in terms of Vasiliev higher-spin theories in $AdS_{d+1}$ \cite{Klebanov:2002ja}. However, this CFT is non-perturbatively unstable due to instanton solutions which give rise to a nonzero imaginary part in the CFT data \cite{Giombi:2019upv}. The Lagrangian reads
 \begin{equation} \label{cubicmodel}
    \mathcal{L} =
  \frac{1}{2} (\partial  \phi_a)^2+ \frac{1}{2} (\partial \eta)^2 - \frac{g}{2}\eta (\phi_a)^2 - \frac{\lambda}{3} \eta^3 \,,
\end{equation}
where $\phi_a$ with $a=1, \dots, N$ is a $O(N)$ vector while $\eta$ is a singlet. For our investigation, we just need to know the one-loop fixed point value of  $\lambda$ which is 
\begin{equation}
      \lambda^* = 3 \sqrt{\frac{6\epsilon(4\pi)^3}{N}}\left(1 + \frac{162}{N}+\frac{68766}{N^2} +... +  \mathcal{O}\left(\epsilon \right)  \right) \,.
\end{equation}
Equipped with the above, we proceed by computing the scaling dimension $\Delta_{n,\eta}$ for the family of composite operators $\eta^n$ in the same double scaling limit considered for the quartic $O(N)$ theory resulting in a semiclassical expansion analogous to Eq. \eqref{dsl}
\begin{equation} \label{lsd}
    \Delta_{n} = n \sum_{i=0} \frac{H_i(\lambda^2 n)}{n^i} \ .
\end{equation}
To this end, we again map our theory on $\mathbb{R}\times S^{d-1}$ and consider a homogeneous field configuration for $\eta$ and a vanishing expectation value for $\phi_a$. The resulting EOM reads
\begin{equation}
    \frac{d^2\eta}{dt^2} + 4 \eta + \lambda \eta^2 = 0 \,,
\end{equation}
and admits the following nontrivial solution
\begin{equation}
    \eta(t)= \frac{1}{\lambda }\left(\frac{6 m \ \text{cn}\left(\left.\frac{t}{\left((m-1) m+1 \right)^{1/4}}\right|m\right)^2-4 m+2}{\sqrt{(m-1) m+1}}-2\right) \,,
\end{equation}
with $0\le m \le 1$. By inserting the above into the following expression for the classical ground state energy
\begin{equation}
   T_{00} = \frac{n}{\pi^3} H_0 = \frac{1}{2} \left(\frac{\partial \eta}{\partial t} \right)^2 + 2\eta^2 +\frac{ \lambda}{3}  \eta^3 \,,
\end{equation}
we obtain the following leading coefficient $H_0$ of the expansion \eqref{lsd}
\begin{equation}
   H_0 =\frac{8 \pi ^3}{3 \lambda ^2 n} \left(\frac{-2 m^3+3 m^2+3 m-2}{((m-1) m+1)^{3/2}}+2\right) \,,
\end{equation}
where $m$ is a nontrivial function of the product $\lambda^2 n$ which is determined by the Bohr-Sommerfeld condition as follows
\begin{equation}
    \frac{2 ((m-1) m+1) \mathcal{E}(m)-(m-2) (m-1) K(m)}{5 ((m-1) m+1)^{5/4}}=\frac{\lambda ^2 n}{48 \pi ^2} \,.
    \label{BSphi3}
\end{equation}
As for the $\phi^4$ theory, $H_0$ resums the leading powers of $n$ at any order of the perturbative expansion for $\Delta_{n}$. Their coefficients can be read off by expanding $H_0$ around $\lambda^2 n =0$. In parallel with our previous analysis, we adopt the notation $H_0 = \sum_{k=0} b_k \left(\frac{\lambda^2 n}{\pi^3} \right)^k$ and provide the first $8$ coefficients below 
\begin{align}
  &  b_0 = 2 \,, \ \ b_1 = -\frac{5}{192} \,,   \ \ b_2  = -\frac{235}{221184} \,,  \nonumber \\ 
  & b_3  = -\frac{38585}{509607936} \,,  \ \  b_4  = -\frac{2663129}{391378894848} \,, \nonumber \\ 
  & b_5 = -\frac{156934505}{225434243432448} \, \ \, b_6 = -\frac{13400341405}{173133498956120064} \,, \nonumber \\  & b_7 = -\frac{7275692993855}{797799163189801254912} \,.
\end{align}
The coefficients $b_0$ and $b_1$ match the results in \cite{Fei:2014yja, Fei:2014xta}. 

Interestingly, the complementary limit of large $\lambda^2 n$ reveals the unstable nature of the theory. In fact, Eq. \eqref{BSphi3} admits real solutions only for $\frac{\lambda^2 n}{(4 \pi)^2} \le \frac65$ where the equality is attained for $m=1$. An analogous behavior has been previously observed in the large charge sector of the theory in \cite{Antipin:2021jiw, Giombi:2020enj} where it has been discovered the existence of a critical value of the charge above which the scaling dimension of the operators carrying the $O(N)$ charges becomes complex. A possible link to instantonic solutions has been later discussed in \cite{Watanabe:2022htq}. In the large $\lambda^2 n$ limit we, therefore, obtain
\begin{equation}
    m \sim e^{\frac{\pm i \pi }{3}}-\left(e^{\frac{\pm i 9 \pi }{10}} 3^{7/10} \left(\frac{16 \pi ^2}{5}  K\left(e^{\frac{\pm i \pi }{3}}\right)\right)^{4/5}\right) \left(\frac{1}{\lambda^2 n}\right)^{4/5} \,,
\end{equation}
leading to
\begin{equation}
    H_0 = \frac{e^{\mp \frac{i \pi}{10}}}{3^{13/10}}  \left(\frac{5 \sqrt{\pi }}{2^{3/2} K\left(e^{\frac{\pm i \pi }{3}}\right)}\right)^{6/5} \left( \lambda^2 n \right)^{1/5} + \mathcal{O}\left(  \left( \lambda^2 n \right)^{-1/5} \right) \,.
\end{equation}
The two complex conjugate solutions correspond to a pair of complex CFTs. We again note the asymptotic  $\Delta_n \sim n^{\frac{d}{d-1}}$ large $n$ behavior previously observed in the large charge sector of the theory \cite{Antipin:2021jiw}. 

 \vskip .2cm
\centerline{\bf The $\phi^6$ theory in $d=3-\epsilon$}
 \vskip .1cm
For our last example, we determine $\Delta_{n}$ for the singlet operators $(\phi_a \phi_a)^{n/2}$ with $a=1, \dots, N$ in the critical $(\phi_a \phi_a)^3$ theory in $d=3-\epsilon$. The Lagrangian reads
\begin{equation}
    \mathcal{L} = \frac{1}{2}(\partial \phi_a)^2-\frac{\lambda^2}{6} (\phi_a \phi_a)^3 \,.
\end{equation}
This theory has a perturbative Wilson-Fisher fixed point whose two-loop value is 
\begin{equation}
   \frac{\lambda^{*2}}{(4 \pi)^2} = \frac{\epsilon}{4(22+3N)} \,,
   \label{FPphi6}
\end{equation}
while the one-loop beta function vanishes identically in $d=3$.
In the same double scaling limit considered for the quartic theory, the scaling dimension of the $\left(\phi_a \phi_a\right)^{n/2}$ operators takes the form of Eq. \eqref{dsl}.
The complete solution of the equation of motion is quite cumbersome and will be treated in a separate work. Here, we limit ourselves to the perturbative small $\lambda n$ regime where the solution of the time-dependent EOM of the model on $\mathbb{R} \times S^{d-1}$ reads
\begin{widetext}  
\small
\begin{align}
   \frac{\phi_a (t)}{ \sqrt{n}} &=\frac{\cos \left(\frac{t}{2}\right)}{\sqrt{\pi }} + \frac{(\lambda  n)^2}{96 \pi ^{5/2}} \left(-60 t \sin \left(\frac{t}{2}\right)-60 \cos \left(\frac{t}{2}\right)+15 \cos \left(\frac{3 t}{2}\right)+\cos \left(\frac{5 t}{2}\right)\right) +\frac{(\lambda  n)^4}{18432 \pi ^{9/2}} \left(-14280 \cos \left(\frac{3 t}{2}\right)  -440 \cos \left(\frac{5 t}{2}\right) \right.  \nonumber \\ & \left.+95 \cos \left(\frac{7 t}{2}\right)+3 \cos \left(\frac{9 t}{2}\right) + 10 \left(\left(4019-360 t^2\right) \cos \left(\frac{t}{2}\right)+3864 t \sin \left(\frac{t}{2}\right)  -60 t \left(9 \sin \left(\frac{3 t}{2}\right)+\sin \left(\frac{5 t}{2}\right)\right)\right)\right) + \mathcal{O}\left( \left( 
 \lambda n\right)^6\right) \,.
\end{align}
\normalsize
\end{widetext}
Replacing the above into the stress-energy tensor one obtains the first few terms of the small $\lambda n$ expansion for $C_0$ 
\begin{equation}
    C_0 =  \frac{1}{2}+\frac{5 \lambda ^2 n^2}{24 \pi ^2}-\frac{131 \lambda ^4 n^4}{384 \pi ^4} + \mathcal{O}\left( \left( 
 \lambda n\right)^6\right) \ .
\end{equation}
To test the solution we insert the fixed point value Eq. \eqref{FPphi6} into the above, and see that it reproduces the leading $n$ term in the $2$-loop expression for $\Delta_n$ which reads \cite{Basu:2015gpa}
\begin{equation}
    \Delta_n = \frac{1}{6(22+3N)} n  (n-2) (5 n+3 N-8) \epsilon \ .
\end{equation}
\vskip .5cm
To summarize our work we have developed a semiclassical framework to compute the scaling dimensions for the class of neutral composite operators in several time-honored CFTs. This has been achieved by considering the double scaling limit  $n\rightarrow \infty$ and $\lambda \rightarrow 0$ with a fixed value of the product $\lambda n$  and employing a saddle point evaluation. We tested our findings at small $\lambda n$  with known diagrammatic results and have been able to predict the infinite series of higher-order terms. Additionally, our leading semiclassical results hold true for various gauge-Yukawa models because fermion and gauge degrees of freedom have vanishing classical backgrounds. Noteworthy examples include the Abelian Higgs and Gross-Neveu-Yukawa models in $d=4-\epsilon$. 

Additionally, our results constitute a strong asset to boost perturbative computations. In fact, one can combine our semiclassical expansion with known perturbative results, at fixed $n$, to determine novel complete higher loop expressions. Importantly, our findings provide an infinite series of checks for future diagrammatic computations.

We plan to go beyond this initial investigation by determining the next semiclassical order $C_1$ stemming from the determinant of the quantum fluctuations around the classical solution. {Therefore, the computations for determining $C_1$ resemble the ones related to computing the leading correction around an instantonic background. As we have shown in this letter, our framework can be extended to several physically relevant quantum field theories in various space-time dimensions.

\vskip .2cm
%\section*{Acknowledgements}
\centerline{\bf Acknowledgements}
 The work of F.S. is
partially supported by the Carlsberg Foundation, semper ardens grant CF22-0922. The work of J.B. was supported by the World Premier International Research Center Initiative (WPI Initiative), MEXT, Japan; and also supported by the JSPS KAKENHI Grant Number JP23K19047. O.A. and J.B. thank the Quantum Theory Center and the Danish Institute for Advanced Study at the University of Southern Denmark for their hospitality and partial support while this work was completed.

\end{document}